\documentclass[aps,prb,twocolumn]{revtex4}
\usepackage{graphicx}
\usepackage{amssymb}

\begin{document}

\title{Calculated NMR $T_2$ relaxation due to vortex vibrations in cuprate superconductors}

\author{Ting Lu and Rachel Wortis}

\affiliation{Department of Physics and Astronomy, Trent University, Peterborough, Ontario,  K9J 7B8, Canada}

\date{\today}

\begin{abstract}
We calculate the rate of transverse relaxation arising from vortex motion in the mixed state of YBa$_2$Cu$_3$O$_7$ with the static field applied along the $c$ axis.  The vortex dynamics are described by an overdamped Langevin equation with a harmonic elastic free energy.  We find that the variation of the relaxation with temperature, average magnetic field, and local field is consistent with experiments; however, the calculated time dependence is different from what has been measured and the value of the rates calculated is roughly two orders of magnitude slower than what is observed.  
Combined with the strong experimental evidence pointing to vortex motion as the dominant mechanism for $T_2$ relaxation, these results call into question a prior conclusion that vortex motion is not significant in $T_1$ measurements in the vortex state.
\end{abstract}


\maketitle

\section{Introduction}
\label{intro}

The motivation for this work is to test a description of vortex motion in the cuprates.  This description, explained below, has been used\cite{wortis:00} to dismiss the importance of vortex dynamics in Cu $T_1$ relaxation in YBa$_2$Cu$_3$O$_7$.  Here it is used to calculate $T_2$ relaxation, widely believed to be dominated by vortex vibrations.  The incomplete agreement between our results and those of experiments has implications for our understanding of vortex dynamics and for conclusions which have been drawn from $T_1$ data.

Many properties of the superconducting state of the cuprates are remarkably conventional, while the normal state remains very poorly understood.  The vortex state in these materials, with coexisting superconducting and normal regions, offers an attractive window through which to examine the normal state as well as excitations of the superconducting state.
NMR has a long history of contributions to our understanding of superconductivity.\cite{parks}
A major strength of NMR is that many different types of measurements can be performed on the same sample with the same apparatus, providing extensive consistency checks of theory.

NMR $T_1$ measurements in the vortex state of high temperature superconductors have been used extensively to study quasiparticle excitations and vortex core states.
\cite{curro:00,mitrovic:03,kakuyanagi:03,takigawa:99,wortis:00,morr:01,knapp:02}
A key advance was the realization\cite{curro:00} that the unique field variation in the vortex lattice state created a strong correspondence between local field and position relative to a vortex, allowing position dependent measurements of relaxation rates.
However, quasiparticles are not the only relaxation mechanism with this position dependence; there are also vortex vibrations. 
An early position averaged calculation\cite{bulaevskii:93} suggested that vortex vibrations were the dominant contributor to Tl $T_1$ relaxation in Tl$_2$Ba$_2$CuO$_6$.
However, another calculation\cite{wortis:00} indicated that the more three-dimensional nature of YBa$_2$Cu$_3$O$_7$ and correspondingly greater stiffness of the vortex lattice meant vortex vibrations made a negligible contribution to Cu $T_1$ in this material.
Both calculations were based on a particular model of vortex motion.
Transverse NMR relaxation provides a forum for evaluating the validity of this model.

$T_2$ measurements in the vortex state have been used to study vortex dynamics near the vortex melting transition, complementing lineshape measurements.\cite{suh:93,borsa:95,borsa:97,recchia:97,nikolaev:97,bachman:98}  A diffusive model of the dynamics has been suggested in this regime.\cite{suh:93}  Our motivation has been to contribute to the understanding of low-energy electronic excitations, which are most easily distinguished at very low temperatures.  We therefore focus on temperatures well below the melting transition.  
Even in this regime, oxygen $T_2$ relaxation in the mixed state of YBa$_2$Cu$_3$O$_7$ is thought to be dominated by vortex vibrations for several reasons.
First, the rates observed at the planar and apical sites are very similar\cite{curro:00} as are those for yttrium\cite{recchia:97}.  
The dominant $T_2$ mechanism for the planar oxygen in the normal state is believed to be dipole coupling with the Cu spins\cite{walstedt:95}.
However, the apical oxygen and the yttrium have a very different crystallographic environment from the planar oxygens and would therefore not be expected to experience the same rate of dipole-induced relaxation, whereas the vortex-motion-induced relaxation could be the same at these different locations due to the continuity of the vortices.
Second, the Cu dipole coupling mechanism gives rise to a Gaussian decay of the spin echo, whereas the echo decays seen in the vortex state are exponential.\cite{bachman:98,curro:00}
Finally, the rate of relaxation varies as a function of local field, being faster at higher local fields and hence closer to the vortex cores\cite{curro:00}, strongly suggesting a vortex related mechanism.

To our knowledge, no analytic calculations of $T_2$ in the vortex state have been done.  Ryu and Stroud\cite{ryu:96} performed numerical Langevin dynamics simulations of the vortex state from which time scales relevant to NMR were extracted.  Experimental results have generally been examined assuming the field-field correlations can be characterized by a single time constant thought to be proportional to the spectral density of the field fluctuations at some low frequency.  The accuracy of this approach has been shown to be poor even in fairly simple models.\cite{witteveen:97}  

We present here (Section \ref{calc}) a calculation of the rate of transverse relaxation due to vortex motion assuming overdamped harmonic vibrations.  The contributions of all modes are included, not just those at a single frequency.  We explore (Section \ref{results}) the time dependence and the local field (position) dependence of the echo decay as well as the temperature and field dependence of the characteristic decay time.  The variation of our results with position, temperature and magnetic field agrees with available data.  However, the time dependence predicted by our calculations is different from that observed, and the absolute value of the rates we calculate are at least two orders of magnitude slower than those seen in experiment.
We discuss (Section \ref{disc}) possible explanations for this discrepancy and implications.
A clear understanding of vortex dynamics is relevant not only to the analysis of $T_1$ measurements as discussed above but also to many fundamental questions such as the nature of vortex core excitations and the significance of Nernst effect measurements\cite{xu:00} as well as applied issues such as efficient enhancement of critical currents.

\section{Calculation}
\label{calc}

In a spin echo measurement,\cite{slichter} the sample sits in a static field ${\bf H}_o=H_o \hat z$ and two pulses of a time varying field ${\bf H}_1$ are applied.  ${\bf H}_1$ is in the $xy$ plane (perpendicular to ${\bf H}_o$) and oscillates at the resonance frequency of a particular nuclear transition.  
The first pulse rotates the magnetization associated with that transition into the $xy$ plane.  
The spins which make up the magnetization are initially all aligned, but due to slight variations in local field which cause slight variations in precession rate, the spins quickly fan out, causing the net magnetization to vanish.
The second ${\bf H}_1$ pulse, applied at a time $\tau$ after the first pulse, rotates the spins 180$^{\circ}$ about an axis $\hat j$ rotating at the average rate of precession.  Slower spins are now ahead and faster ones behind, and at a time $2\tau$ after the initial ${\bf H}_1$ pulse the spins realign, causing a peak in the net magnetization.
The transverse relaxation time, $T_2$, is the time scale for the decay of this spin echo with delay time $\tau$ and represents dephasing due specifically to time-dependent local field variations.  The effect of static field variations is removed in the spin echo process.  Longitudinal ($T_1$) relaxation contributes in principle to the spin echo decay; however, $T_1$ is generally much greater than $T_2$ and hence this effect can be neglected.  ($T_1 \sim 1000 T_2$ for planar oxygen in YBCO.\cite{curro:00})

We calculate here the magnitude of the net magnetization in the plane perpendicular to ${\bf H}_o$ at the time of the spin echo, $2\tau$.  
Our calculation is made possible by assuming that the probability distribution of the angle through which a given spin has precessed, $\phi$, is Gaussian.\cite{recchia:96}  If this is the case,
\begin{eqnarray}
M(2\tau)
&=&
M_o e^{-\langle \phi^2(2\tau)\rangle/2}
\end{eqnarray}
where $M_o$ is the initial magnetization and $\langle \phi^2(2\tau) \rangle$ is the thermal average of the square of the precession angle at the time 2 $\tau$.  Because the instantaneous precession frequencies are continuous in time, this Gaussian approximation is expected to be fairly accurate.\cite{recchia:96}
$\phi$ is simply proportional to the time integral of the local field.  Taking the effect of the second ${\bf H}_1$ pulse into account (and assuming its duration is negligible), the second moment of the precession angle may be written in terms of the field-field correlation function:
\begin{eqnarray}
\langle \phi^2 (2\tau) \rangle 
&=&
\gamma_n^2
\left( \int_{0}^{\tau} - \int_{\tau}^{2\tau}  \right)^2
d t \ d t' \langle h_z(t) h_z(t') \rangle.
\label{phi2}
\end{eqnarray}
where $\gamma_n$ is the nuclear gyromagnetic ratio.

The rest of the calculation relies on a description of the dynamics of the local field for which we have made the following fairly standard\cite{brandt:95,clem:92,ryu:96} assumptions. 
First, because all measurements to date have been done on YBCO which has moderate anisotropy, we use a three-dimensional anisotropic continuous model rather than a Lawrence-Doniach description of the superconductor.  
We keep $c$-axis versus plane anisotropy, but neglect $ab$ anisotropy.
We consider a static field ($H_o$) applied parallel to the $c$ axis and assume the equilibrium positions of the vortices form a hexagonal lattice.  
The vortex field profile is calculated from London's equation with a Gaussian short-distance cutoff.
\begin{eqnarray}
h_z({\bf r},t) &=& \int {d{\bf p} \over (2 \pi)^3} e^{+i{\bf p}\cdot {\bf r}} h_z({\bf p},t) \\
h_z({\bf p},t) &=& \Phi_o \sum_i \int dz_i 
{e^{-i{\bf p}\cdot{\bf r}_i(t)} e^{-p_{\perp}^2 \xi_{ab}^2/4 - p_z^2 \xi_c^2/4} \over (1 + \lambda_{ab}^2p^2)} \\
{\bf r}_i(t) &=& {\bf r}_i^o + {\bf u}_i(z_i,t) 
\\
{\bf u}_i(z_i,t) &=& \int {d{\bf k} \over (2\pi)^3} e^{+i{\bf k}\cdot{\bf r}_i^o}
\left[ u_{\ell}({\bf k},t) \hat k_{\perp} + u_t({\bf k},t) \hat z \times \hat k_{\perp} \right]
\end{eqnarray}
${\bf r}_i^o$ is the equilibrium position, and ${\bf u}_i$ is the horizontal displacement of vortex $i$.
$u_{\ell}$ and $u_t$ are the longitudinal and transverse Fourier components of the displacement.
$\lambda_{ab}$ and $\xi_{ab}$ are the magnetic penetration depth and the superconducting correlation length in the $ab$ plane.  $\xi_c = \xi_{ab}/\Gamma$ and $\lambda_c = \Gamma \lambda_{ab}$.

The vortex dynamics are described by a Langevin equation:  
Each vortex segment (length $d$) feels an elastic restoring force given by the derivative of the vortex lattice free energy,
a viscous force proportional to its velocity with constant of proportionality $\eta$, and a random thermal noise term.
\begin{eqnarray}
& &
-{\delta {\cal F}_{elastic} \over \delta {\bf u}_i}
-\eta d {\partial {\bf u}_i \over \partial t} 
+ d {\vec \xi}_i
= 0
\\
& &
{\cal F}_{elastic} = {1 \over 2} \int {d{\bf k} \over (2\pi)^3} 
\Biggl\{ \epsilon_{\ell}(k_{\perp},k_z) |u_{\ell}({\bf k},t)|^2 
\nonumber \\
& & \hskip 1.5 in + \epsilon_t(k_{\perp},k_z) |u_t({\bf k},t)|^2 \Biggr\} 
\end{eqnarray}
The inertial term is neglected.
Details of the vortex lattice elastic properties are described in a review by E. Brandt.\cite{brandt:95}
From these come the displacement-displacement correlations.
\begin{eqnarray}
\langle u_{\ell}({\bf q},t) u_{\ell}^*({\bf q},t') \rangle
&=& {V k_B T \over \epsilon_{\ell}(q_{\perp},q_z)} e^{- \gamma_{\ell}({\bf q}) |t-t'|} \\
{\rm where} \ \gamma_{\ell}(q_{\perp},q_z) &=& {\Phi_o \epsilon_{\ell}(q_{\perp},q_z) \over B \eta}
\end{eqnarray} 
where $B$ is the average internal field and $\Phi_o$ is the flux quantum, 
and similarly for the transverse correlations.

In calculating the field-field correlation function, we sum over equivalent vortex lattice positions, $i$, but retain position dependence within the unit cell, $\vec \rho$.
\begin{eqnarray}
G({\bf r},t-t') &\equiv& \langle h_z({\bf r},t) h_z({\bf r},t') \rangle \\
G({\vec \rho},t-t')
&=& {1 \over N_{\perp}} \sum_i {1 \over L_z} \int dz G({\bf r},t-t') \nonumber \\
&=&
{1 \over A_{\perp} L_z}
\sum_{{\bf g}} \sum_{{\bf g}'} e^{+i{\bf g}\cdot{\vec \rho}} 
\int_{BZ} {d {\bf k} \over (2 \pi)^3} 
\nonumber \\
& & \times
\langle h_z({\bf g}'+{\bf k},t) h_z({\bf g}'+{\bf k}-{\bf g},t') \rangle 
\end{eqnarray}
Here ${\bf g}$ and ${\bf g}'$ are vortex lattice reciprocal lattice vectors, and ${\bf k}$ is limited to the first Brillouin zone.
$\langle h_z({\bf g}'+{\bf k},t) h_z({\bf g}'+{\bf k}-{\bf g},t') \rangle$ can be written in terms of the longitudinal and transverse displacement-displacement correlations using the harmonic form of the elastic free energy to write 
\begin{eqnarray}
& & \langle e^{-i({\bf g}'+{\bf k})\cdot{\bf r}_i(t)}e^{-i({\bf g}-{\bf g}'-{\bf k})\cdot{\bf r}_j(t')} \rangle
\nonumber \\
& & =
e^{-i{\bf k}\cdot{\bf r}_i^o} e^{+i{\bf k}\cdot{\bf r}_j^o}
e^{-{1 \over 2}\langle [({\bf g}'+{\bf k})\cdot{\bf u}_i(z_i,t)]^2 \rangle}
\nonumber \\
& & \hskip 0.3 in \times
e^{-{1 \over 2}\langle [({\bf g}-{\bf g}'-{\bf k})\cdot{\bf u}_j(z_j,t')]^2 \rangle}
\nonumber \\
& & \hskip 0.3 in \times
e^{- \langle [({\bf g}'+{\bf k})\cdot{\bf u}_i(z_i,t)][({\bf g}-{\bf g}'-{\bf k})\cdot{\bf u}_j(z_j,t')] \rangle}
\label{5exps}
\end{eqnarray}
The Debye-Waller factors may be simplified by neglecting the hexagonal anisotropy of the vortex motions:
\begin{eqnarray}
e^{-{1 \over 2}\langle [({\bf g}'+{\bf k})\cdot{\bf u}_i(z_i,t)]^2 \rangle}
&\sim&
e^{-{1 \over 4} |{\bf g}'+{\bf k}|^2 \langle u^2 \rangle}
\end{eqnarray}
where $\langle u^2 \rangle$ is the mean square deviation of a vortex from its equilibrium position.
We expand $e^{- \langle [({\bf g}'+{\bf k})\cdot{\bf u}_i(z_i,t)][({\bf g}-{\bf g}'-{\bf k})\cdot{\bf u}_j(z_j,t')] \rangle} \sim 1 - \langle [({\bf g}'+{\bf k})\cdot{\bf u}_i(z_i,t)][({\bf g}-{\bf g}'-{\bf k})\cdot{\bf u}_j(z_j,t')] \rangle$.  The zeroth-order term is independent of time and hence gives zero when the time integral is performed, leaving only the linear term.  This appears to be analogous to the one-phonon approximation from neutron scattering.\cite{ashmer}  However, because the physical meaning--and magnitude--of the quantities involved are not directly analogous, we have verified the approximation numerically.\cite{lu:05}

Using displacement-displacement correlations derived from the Langevin dynamics 
and performing the time integral, we obtain the following expression for the second moment of the phase distibution:
\begin{eqnarray}
& &
\langle \phi^2 (\vec \rho, 2 \tau) \rangle
\nonumber \\
& & =
{2 \gamma_n^2 k_B T B \Phi_o \tau^3 \over \eta}
\sum_{\bf g} e^{+i {\bf g}\cdot{\vec \rho}}
\sum_{{\bf g}'}
\int_{BZ} {d{\bf k} \over (2 \pi)^3}
\nonumber \\
& & \hskip 0.2 in
\times
{e^{-|{\bf g}'+{\bf k}_{\perp}|^2 \xi_{ab}^2/4 - k_z^2 \xi_c^2/4}
\over (1 + \lambda_{ab}^2 |{\bf g}' + {\bf k}|^2)}
{e^{-|{\bf g}'+{\bf k}_{\perp}-{\bf g}|^2 \xi_{ab}^2/4 - k_z^2 \xi_c^2/4}
\over (1 + \lambda_{ab}^2 |{\bf g}'+{\bf k}-{\bf g}|^2)} 
\nonumber \\
& & \hskip 0.2 in
\times
e^{-|{\bf g}' + {\bf k}_{\perp}|^2 \langle u^2 \rangle/4}
e^{-|{\bf g}' + {\bf k}_{\perp} - {\bf g}|^2 \langle u^2 \rangle/4}
\nonumber \\
& & \hskip 0.2 in
\times
\Biggl\{
\left[ ({\bf g}' + {\bf k}_{\perp})\cdot{\hat k_{\perp}} \right]
\left[ ({\bf g}' + {\bf k}_{\perp}- {\bf g})\cdot{\hat k_{\perp}} \right]
\nonumber \\
& & \hskip 0.4 in
\times
{\left(- 3 + 2 \gamma_{\ell} \tau + 4 e^{- \gamma_{\ell} \tau}
                                  - e^{- 2 \gamma_{\ell} \tau} \right)
 \over (\gamma_{\ell} \tau)^3}
\nonumber \\
& & \hskip 0.4 in
+
\left[ ({\bf g}' + {\bf k}_{\perp})
       \cdot{\hat z \times \hat k_{\perp}} \right]
\left[ ({\bf g}' + {\bf k}_{\perp}- {\bf g})
       \cdot{\hat z \times \hat k_{\perp}} \right]
\nonumber \\
& & \hskip 0.6 in
\times
{\left(- 3 + 2 \gamma_{t} \tau + 4 e^{- \gamma_{t} \tau}
                                 - e^{- 2 \gamma_{t} \tau} \right)
 \over (\gamma_{t} \tau)^3}
\Biggr\}
\end{eqnarray}

Unlike in more strongly layered materials,\cite{bulaevskii:93} this cannot be significantly simplified by physically reasonable approximations.  We have therefore calculated the sums numerically.  An adaptive step size approach was necessary in order to obtain reliable results in a reasonable amount of time given the very fast variation at small ${\bf k}$ values and slow variation at higher momenta with a time dependent crossover between the two regimes.  For ease of calculation, a rectangular unit cell was used in ${\bf k}$-space.  The symmetry of our results is unaffected by this approximation, implying that very high ${\bf k}$ values, where our harmonic approximation is least justified, do not play a key role.  Each sum on ${\bf g}$ and ${\bf g}'$ includes all reciprocal lattice vectors which make a significant contribution.  This means magnitudes of ${\bf g}$ up to roughly $2 \pi/\xi_{ab}$, corresponding to the short-distance cutoff.  Values of ${\bf g}'$ between ${\bf g}$ and the origin dominate, and concentric rings sufficient to obtain convergence are included.

The model includes two externally controlled parameters, temperature and field, for which we used values consistent with available experiments: temperatures between 10 and 50 K and average internal fields between 5 and 25 T.  
There are five material parameters required, for which we used the following values.
The gyromagnetic ratio of $^{17}$O, $^{17}\gamma \sim 3.627 \times 10^{7}$ 1/sT.
For the in plane magnetic penetration depth and superconducting correlation length of YBCO we used 1600 \AA \ and 16 \AA \ respectively, with an anisotropy ratio $\Gamma$ of 5.
The vortex viscosity coefficient, $\eta$, is inferred from complex surface impedance measurements to be of order 10$^{-6}$ kg/ms.\cite{tsuchiya:01}
From these parameters, the mean square deviation of the vortices, $\langle u^2 \rangle$ may be calculated.\cite{brandt:95}
As a check on our model and parameters, we calculated $\langle u^2 \rangle$ at temperatures and fields where vortex melting is observed and found values consistent with the Lindemann criterion.

\section{Results}
\label{results}

Two main factors affect the rate of relaxation. 
First, the larger the gradients in the local field, the faster dephasing will occur, simply because it is the field variation which gives rise to a range of precession frequencies.
Second, the closer the match between the characteristic frequencies of the vortices, $\gamma_{\ell}$ and $\gamma_t$, and NMR time scales the shorter $T_2$.
If the vortices move very slowly the field gradient is effectively static and hence does not cause decay of the spin echo.
If the vortices move very quickly, each nucleus experiences effectively an average field, and again there is no time dependence.
Our calculation suggests that the motions are on the fast side of the maximum between these two limits.  Contributions from the transverse modes are roughly two orders of magnitude greater than those from the longitudinal modes.  The transverse modes are softer and hence correspond to lower frequencies.
Likewise, although all $k$ values contribute, it is the small $k$ (long wavelength) modes which dominate at long times.

\begin{figure}
\includegraphics[angle=-90,width=\columnwidth]{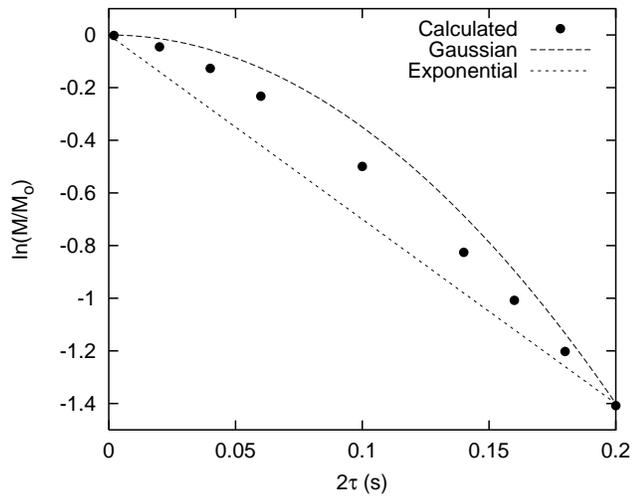}
\caption{Calculated echo decay at $B=$10 T and $T=$40 K using model and parameters described in text. Gaussian and single exponential lines are shown for comparison.  \label{taudep}}
\end{figure}

Figure \ref{taudep} shows the spin echo height as a function of time $2\tau$ following the initial ${\bf H}_1$ pulse.  
The results shown were calculated at 40 K and an average internal field of 10 T and are representative of the qualitative behavior seen throughout the temperature and field range we examined.
The decay is neither Gaussian, as in the Cu $T_1$ mechanism\cite{walstedt:95}, nor simple exponential, as seen in measurements in the vortex state\cite{bachman:98}.  

\begin{figure}
\includegraphics[angle=-90,width=\columnwidth]{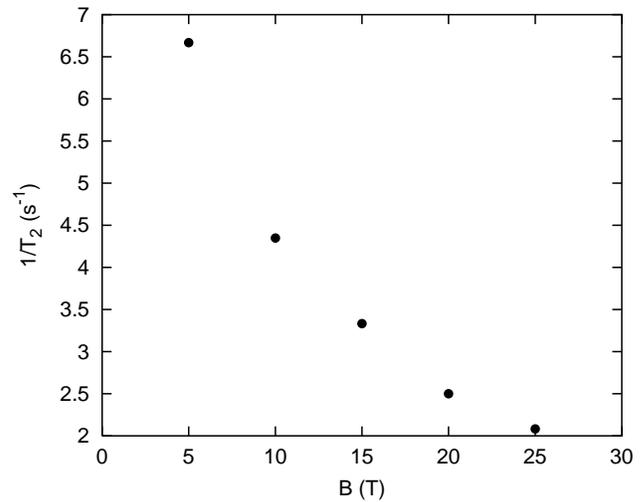}
\caption{Calculated transverse relaxation rate as a function of average mangetic field, $B$, at $T=$ 20 K. \label{Bdep}}
\end{figure}

We define $T_2$ to be the time at which the echo decays to $1/e$ its initial value.
Figure \ref{Bdep} plots $1/T_2$ as a function of average internal field at a temperature of 20 K.  The rate decreases with increasing magnetic field.  This represents a combination of two effects:  First, the field distribution becomes somewhat smoother at higher fields causing less local field variation for the same magnitude of vortex motion.  Second, the vortex lattice is stiffer at higher fields causing a greater mismatch between the characteristic vortex frequencies and the NMR time scale.  A decline in the rate of exponential echo decay with increasing magnetic field was observed by the Bachman, et al\cite{bachman:98}.  

\begin{figure}
\includegraphics[angle=-90,width=\columnwidth]{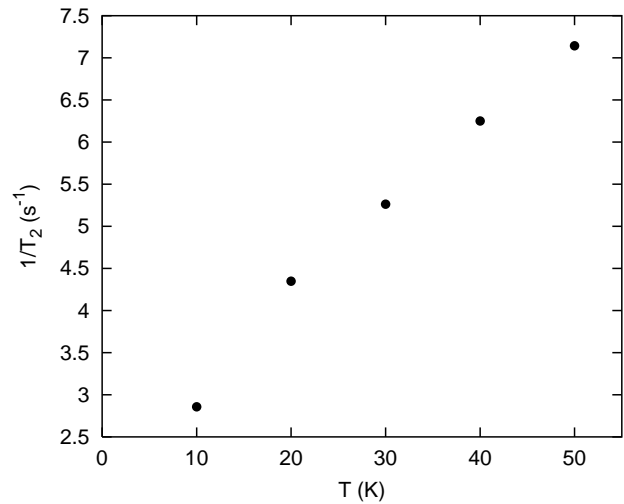}
\caption{Calculated relaxation rate as a function of temperature at an average field $B=$ 10 T. \label{Tdep}}
\end{figure}

Figure \ref{Tdep} plots $1/T_2$ as a function of temperature at an average internal field of 10 T.  The rate increases essentially linearly with temperature.  This represents greater occupation of vibrational modes at higher temperatures, i.e. larger amplitude motions.  Curro, et al\cite{curro:00} show a linear increase in the transverse relaxation rate--interrupted at 30 K by a peak which appears to be field independent and hence not vortex related.

\begin{figure}
\includegraphics[angle=-90,width=\columnwidth]{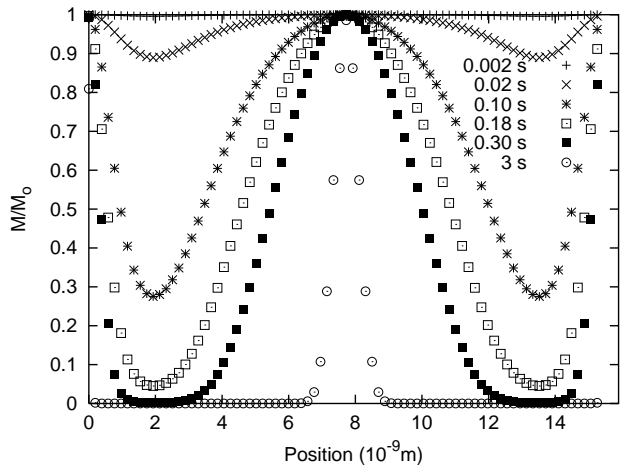}
\caption{Calculated echo magnitude as a function of position between two neighboring vortices at $B=$10 T, $T=$20 K and a series of times, $2\tau$, shown in key. \label{rdep}}
\end{figure}

All the results presented above represent averages over the whole sample.  Figure \ref{rdep} shows the height of the spin echo as a function of position along the line between two vortices at 20 K, an average internal field of 10 T and a series of delay times.  At the center of each vortex and at the saddle point between them, the field gradient goes to zero and the relaxation is vanishingly small.  At a distance just over $\xi_{ab}$ from the center of each vortex, where the field gradient is very large, the relaxation is fastest.

\begin{figure}
\includegraphics[angle=-90,width=\columnwidth]{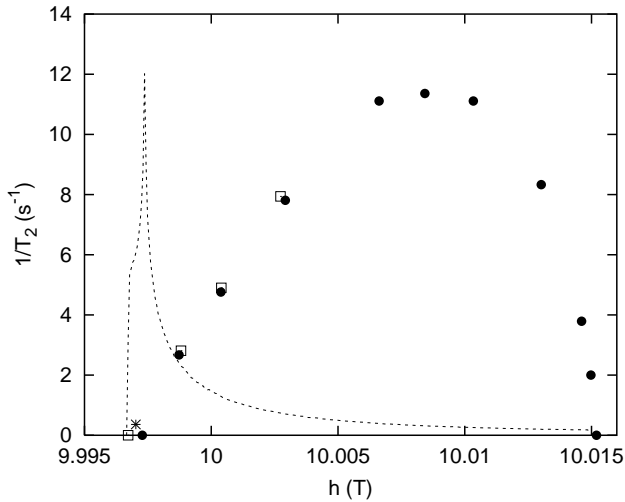}
\caption{Points show calculated transverse relaxation rate as a function of local field.   Filled circles correspond to positions on a line between neighboring vortices.  Open squares correspond to positions along a line passing through a vortex and an adjacent field minimum.  The star is a point half way between the field minimum and the saddle point.  The dashed line sketches the theoretical vortex lattice resonance lineshape.  \label{hdep}}
\end{figure}

NMR measurements cannot look directly at this position dependence, but a strong correlation can be made between position and local field.  Figure \ref{hdep} shows $1/T_2$ as a function of local field, again at 20 K and an average internal field of 10 T.
The dashed line sketches the theoretical resonance lineshape.  Note that there can be multiple rates corresponding to a single local field.  This is both because the rate depends on the field gradient and not the field itself and because of the hexagonal rather than circular symmetry of the vortex unit cell.  The range of rates at a given field is, however, small.

Curro, et al\cite{curro:00} measured $1/T_2$ as a function of position in their significantly broadened resonance line.  Their results show a rise by a factor in the range 1.5 to 2 between the lowest and highest local fields.  
The broadening of their resonance line implies an averaging of adjacent rates, which is consistent with their having measured nonzero $1/T_2$ at the low field end.  Furthermore, the loss of signal at high fields would have made it difficult to see the downturn in the high field rates, although some temperatures show leveling off at high fields.

\section{Discussion}
\label{disc}

The trends we calculate are consistent with the temperature, average field and local field (i.e. position) dependence seen in experiments.  
However, the time dependence of the echo decay is different from that which is observed, and there is a major mismatch between the rates we calculate  (of order 10 1/s) and the rates seen in experiments (of order 1000 1/s).
A mismatch of this magnitude was noted earlier by experimentalists\cite{curro:00,recchia:97} comparing single mode models with numerical calculations\cite{ryu:96}.
We have shown that this discrepancy is inherent to the overdamped vortex vibration model and not simply created by the approximations involved in the single-mode comparison. 
Possible explanations for the discrepancy include (i)  that the physical parameters we use are inaccurate, (ii) that our model of vortex dynamics is inappropriate, and (iii) that vortices are not in fact the dominant mechanism for transverse relaxation.
The third possibility seems especially unlikely for all the reasons given in the introduction.

Regarding the first possibility, we have studied the sensitivity of our results to the four material parameters $\Gamma$, $\xi_{ab}$, $\lambda_{ab}$, and $\eta$.
The anisotropy enters primarily through the vortex lattice elastic constants\cite{brandt:95}, and the dependence is extremely weak:
doubling $\Gamma$ changes the rate by less than 5\%.
The primary role of the correlation length is in determining the field distribution near the core.  
Smaller values of $\xi_{ab}$ mean larger field gradients and hence faster relaxation.  However, cutting the correlation length from 16 \AA \ to 10 \AA \ reduces $T_2$ by roughly 20\%.  
The magnetic penetration depth influences both the field gradient and the stiffness of the lattice.  
A larger $\lambda_{ab}$ produces less variation in the local field and a softer vortex lattice.
Of these two competing effects, the first is stronger.  Reducing $\lambda_{ab}$ from 1600 \AA \ to 1000 \AA \ reduces $T_2$ by almost 50\%.
Finally, the vortex viscosity influences the characteristic frequencies.  As discussed above, both very high frequencies (small $\eta$) and very low frequencies (large $\eta$) produce slow relaxation.  
An increase in $\eta$ from 10$^{-6}$ kg/ms to 10$^{-5}$ kg/ms slows the vortices and reduces $T_2$ by about 50\%.
Even with all of these changes combined, our calculated rate is at best still more than an order of magnitude slower than that observed.

Regarding the possibility that the vortex dynamics are not captured by our model, the overdamped Langevin dynamics picture is widely used.\cite{brandt:95,clem:92,ryu:96}
Undamped motion was explored in the context of organic superconductors,\cite{xing:94} but would produce even higher characteristic frequencies.
Another possibility might be a slower motion superimposed on the faster motion we've described, for example flux creep of vortex bundles.  However, stimulated echo measurements\cite{recchia:97} suggest that the vortex motions are small compared to their lattice spacing.
As for the influence of disorder, the average intervortex spacing is not dramatically changed \cite{shibata:05} by disorder and our results appear not to be very sensitive to short wavelength disturbances.

A key feature of our implementation of this model, however, appears to be the assumption of a continuum of vibrational modes, corresponding to an infinite system size.
The influence this has on the time dependence can be seen as follows.
For each mode, when $\gamma \tau << 1$ $\langle \phi^2 \rangle \propto \tau^3$ and when $\gamma \tau >> 1$ $\langle \phi^2 \rangle \propto \tau$.  Because in our model there is a continuum of modes down to zero energy, one never reaches the long time limit of all modes and therefore there is always curvature.
Available vibrational modes will be restricted in a finite size system, such as the powder samples used in NMR, and also in a system in which vortex pinning is significant.
A very simple model in which short wavelength modes are simply removed does indeed produce an exponential echo decay as seen in experiments.
However, a more careful exploration of the effect of a discrete spectrum on our results is still in progress.

In conclusion, we have calculated the time, temperature, average field, and local field (position) dependence of the rate of transverse NMR relaxation arising from vortex motion using an overdamped Langevin dynamics model and harmonic elastic free energy.  The functional dependence of our results on temperature, field and position are consistent with available experiments, while the time dependence and the rate itself are not.
A key implication of this result is that vortex motion cannot necessarily be neglected as a mechanism for $T_1$ relaxation in the vortex state, influencing conclusions which have been drawn from these measurements on the nature of the electronic states in the presence of vortices.

\begin{acknowledgements}
We gratefully acknowledge John Berlinsky and Catherine Kallin for their contributions to the early stages of this work.  We also thank Cristina Chifor and Will Darling for their computational efforts, and Bill Atkinson for helpful conversations and numerical advice.  
This work has been supported by the Natural Sciences and Engineering Research Council of Canada.
\end{acknowledgements}


\end{document}